\documentclass[amsmath,amssymb,aps,twocolumn]{revtex4-2}
\usepackage{graphicx}
\usepackage{dcolumn}
\usepackage{bm}
\usepackage{xcolor}
\begin{document}
\title{Turbulence: An entropic approach} %
\author{Christian Beck$\,^{1}$}
\email{c.beck@qmul.ac.uk}
\author{Constantino Tsallis$\,^{2,3,4,5}$}
\affiliation{$\,^1$Centre for Complex Systems, Queen Mary University of London, School of Mathematical Sciences, Mile End Road, London E1 4NS, UK}
\affiliation{$\,^2$Centro Brasileiro de Pesquisas F\'{\i}sicas, Rua Xavier Sigaud 150, 22290-180, Rio de Janeiro RJ, Brazil}
\affiliation{$\,^3$Santa Fe Institute, 1399 Hyde Park Road, Santa Fe, 87501 NM, USA}
\affiliation{$\,^4$Complexity Science Hub Vienna,  
Metternichgasse 8, 1030 Vienna, Austria}
\affiliation{$\,^5${Dipartimento di Fisica e Astronomia Ettore Majorana, University of Catania, Via S. Sofia 64, Catania 95123, Italy}}

\begin{abstract}
{We point out a new physical application for non-additive entropic functionals $S_{q,\delta}$ in turbulence.
Maximizing $S_{q,\delta}$ subject to standard 
kinetic energy constraints} provides generalized canonical distributions that agree perfectly with measured probability densities of velocity differences at distance $r$ in highly-turbulent Taylor-Couette flow. The end point of the turbulent cascade is described by $\delta =\frac{3}{2}$, a parameter value that also plays an important role in black-hole physics. At this point the Kolmogorov length scale $r=\eta$ is reached and all observable eddy structures of the turbulent flow disappear, in certain analogy to what is observed for black holes at the event horizon. Our approach generalizes statistical mechanics to 
nonadditive  entropic functionals $S_{q,\delta}$ such that it is applicable to turbulent flows. This approach asymptotically generates stretched $q$-exponentials as generalized canonical distributions relevant for turbulent flow, with a particular dependence of the stretching exponent $\delta^{-1}$ on $q$ that follows from the well-known escort formalism in nonextensive statistical mechanics. Along this particular line in the parameter space, the physics can be described by $S_q$ on its own with suitable escort constraints, leading to the prediction $\delta^{-1} (r) =2-q(r)$, thus allowing for a consistent thermodynamic description since $S_q$ is both trace-form and composable. We show that the above theoretically derived relation is well satisfied by measured 
high-precision experimental data for Taylor-Couette flow. At the Kolmogorov length scale $r=\eta$, the endpoint of our scenario, one has $\delta =\frac{3}{2}$ and  at this point the third moment of velocity differences ceases to exist and all eddies disappear. 
\end{abstract}

\maketitle


Hydrodynamic turbulence can be regarded as a fundamental scientific challenge which has not been completely solved so far \cite{sreeni2025}. Among several open problems, a very interesting 
one is the understanding of the scale-dependent statistics of radial velocity differences $u$ measured in a fully developed turbulent flow at distance $r$. The corresponding probability density functions (PDFs) are non-Gaussian and exhibit complicated behaviour as a function of the scale $r$, the Reynolds number $Re$, and the boundary conditions. This complexity translates into a corresponding complexity for the moment scaling exponents and other important quantities characterizing the turbulent flow.

While direct numerical simulations (DNS) of the Navier-Stokes equation can of course be done and lead to good agreement with many experimental results, a deeper understanding of the scaling properties of the PDFs is still missing.
A simplified theoretical approach is based on a variety of turbulent cascade models (often with multifractal properties) that are in reasonably good agreement with some measured moment scaling exponents, see, e.g.,  \cite{benzi,sreeni2,ruelle}. One can also implement simple dynamical mixture models
associated with these types of cascades to model the PDFs involved
\cite{castaing, beck1, beck2}.
However, a deeper understanding why such simplified models are applicable and sometimes yield good results in agreement with experiments is still missing. Moreover, the intermittency parameters in these types of models need to be fitted, {they cannot be directly derived from the Navier-Stokes equation \cite{openai2026}}. For recent work dealing with DNS data of the forward and inverse energy cascade in fluid turbulence, as well as the circulation statistics, see \cite{yao2024,sreeni2026}. 
{These recent papers, as well as earlier work such as  \cite{beck1, beck2,BCS, ruelle, BLS} have made it clear that methods borrowed from nonequilibrium statistical physics are relevant and useful for the effective description of turbulent flow statistics.}

In this Letter we provide a new theoretical approach to turbulence that is based on a novel entropic (information-theoretic) formalism. {Our aim is to obtain a better understanding of the scaling properties of PDFs of velocity differences at a given scale $r$. We will compare with experimental data previously discussed in \cite{BLS}. Our  novel theoretical approach is based on maximizing two-parameter entropic forms $S_{q,\delta}$. These entropic functional forms were previously introduced in \cite{Tsallis-Cirto} in a black hole context, but the application to turbulent flows presented here is new.} Basically, we develop a new (generalized) statistical mechanics formalism that is applicable (and experimentally testable) for turbulent flows. {The advantage of our approach is that it is using ordinary kinetic energy constraints
rather than introducing artificial stretching.}
Technically, this method is based on the maximization of the generalized entropy measures $S_{q,\delta}$  (depending on two parameters $q$ and $\delta$) subject to suitable energy constraints, {which represent the fluctuating kinetic energy of the eddies in the turbulent flow at a given scale}. As we will show, this new approach is very successful to quantitatively understand the probability distributions of stationary states in the space of scale-dependent velocity differences. We will compare with {a benchmark set of} measured precision data from turbulent Taylor-Couette flow at high Reynolds numbers \cite{BLS}, obtaining excellent agreement.

Our theory gives a concrete prediction on the relation between $q$ and $\delta$ that turns out to be satisfied by experimentally measured turbulent Taylor-Couette flow data, thus confirming the applicability of our new theory. 
This theory opens up a new application of generalized statistical mechanics to understand turbulence from an information-theoretic statistical physics point of view, giving concrete theoretical predictions for parameter dependencies that, as we will show, are confirmed by experiments. 

The generalized entropy $S_{q,\delta}$ that is suitable for turbulent flow descriptions has been introduced in a different context in \cite{Tsallis-Cirto} and it has recently received strong attention when applying the $\delta$-dependent part to cosmological data of the universe as a whole \cite{jizba2024, luciano2025, TsallisJensen2025, jizba2022}. The special value $\delta=\frac{3}{2}$ and $q=1$ corresponds to a thermodynamic black hole entropy. We show that this particular $\delta$-value $\delta =\frac{3}{2}$ corresponds to the end point of the cascade in our turbulence model, where all turbulent eddy structures fully dissipate at the Kolmogorov scale
(in a certain formal analogy to a black hole that is absorbing all turbulent structures).

In the final part of our paper we will take this  analogy a step further and ask if our information-theoretic turbulence theory could potentially be applied to the turbulent surroundings of real (astrophysical) black holes. Rapidly spinning black holes are expected to  trigger turbulent flows of matter in their surroundings \cite{Yang-et-al}, 
and in fact turbulent behaviour has been recently reported for the vicinity of supermassive black holes at the centre of the galaxy \cite{nature2026,xrism2025}. Naturally, as
black holes are very well-described by information-theoretic thermodynamic approaches, and turbulence is as well (as shown in this Letter), the two approaches can be combined. 


{\it Maximizing $S_{q,\delta}$ -} Our theory in the following is based on maximizing the generalized 2-parameter entropic functional $S_{q,\delta}$ subject to a normalization and energy constraint. We have \cite{Tsallis-Cirto}
\begin{equation}
    S_{q,\delta}=\sum_{i=1}^Wp_i\left( \ln_q \frac{1}{p_i} \right)^\delta
\end{equation}
where the $q$-logarithm is defined as
\begin{equation}
    \ln_q x := \frac{x^{1-q}-1}{1-q}
\end{equation}
and the $p_i$ denote the probabilities of the microstates $i$.
Note that
\begin{equation}
    \ln_{2-q} x = \frac{x^{-1+q}-1}{-1+q} =-\ln_q \frac{1}{x}
\end{equation}
In the following we assume that $q<1$ such that $q_{ss}=2-q>1$. The value $q_{ss}$ will be the relevant index to describe the {\it stationary state}, i.e.\ the generalized canonical distributions exhibiting power law decay that we will later compare with
experimentally measured turbulence data as reported in \cite{BLS}.
The relation $q_{ss}=2-q$ also occurs for
other complex systems \cite{prl-andrade,dogninitsallis}.

Following similar lines as in \cite{Tsallis1988, Tsallis2023book, Beck-review} we consider the
free energy functional
\begin{equation}
    \Psi [p] =- \sum p_i \left( \ln_q \frac{1}{p_i}\right)^\delta + \alpha \sum p_i + \beta \sum p_i E_i
\end{equation}
which should be minimized as a function of the set of all the $p_i$, denoted by $[p]$. The extremum condition
\begin{equation}
    \frac{\partial}{\partial p_i}\Psi [p]=0\;\;\;\;\;\forall i
\end{equation}
leads, after a short calculation, to the equation 
\begin{equation}
    \left( \ln_q \frac{1}{p_i} \right)^\delta \left( 1- \delta \cdot p_i^{q-1} \left( \ln_q \frac{1}{p_i}\right)^{-1} \right)= \alpha +\beta E_i \,.
    \label{6}
\end{equation}
The correction term proportional to $\delta$ in the 2nd bracket is small if $p_i$ is small (meaning we are looking at the tails), and this term generally has only a very weak dependence on $p_i$. {In fact,
we can write the correction term as
\begin{equation}
    \delta \cdot p_i^{q-1} \frac{1}{\ln_q \frac{1}{p_i}}=\delta \cdot p_i^{q-1} \frac{1-q}{\left(\frac{1}{p_i}\right)^{1-q}-1}  =h(p_i) \delta (1-q) 
\end{equation}
where the function $h(p_i)$ has only a very weak dependence on $p_i$ in the range of parameters that is relevant for experimental turbulence data. For example, choosing the two very different values $p_i=10^{-2}$ and $p_i=10^{-4}$, which differ by a factor 100,  $h(p_i)$ takes on the two very similar values 0.773 and 0.880 for $1-q=0.02$, respectively 0.608 and 0.729 for $1-q=0.18$, i.e.\ there is only a very small variation of $h(p_i)$ by less than 20\% if $p_i$ is changed by a factor 100. This extremely slow
variation of $h(p_i)$ as a function of $p_i$ is not visible in a logarithmic plot when comparing with turbulence data, in particular when all distributions are finally normalized. 
Thus, in very good approximation we may use $h(p_i)\approx const$. With this Eq.~(\ref{6}) implies }
\begin{equation}
    \left( \ln_q \frac{1}{p_i} \right)^\delta =\alpha + \beta E_i
\end{equation}
where the parameters $\alpha$ and $\beta$ have been  both rescaled by the constant factor $1/(1-const \cdot \delta (q-1))$ and then, to simplify the notation, re-named into $\alpha$ and $\beta$. The above can be written as
\begin{equation}
    \ln_q \frac{1}{p_i} =(\alpha +\beta E_i)^\frac{1}{\delta}
\end{equation}
which is equivalent to
\begin{equation}
    p_i=\left( 1+(1-q) (\alpha +\beta E_i)^\frac{1}{\delta}\right)^\frac{1}{q-1}. \label{10}
\end{equation}
Using again
\begin{equation}
    q_{ss}=2-q>1
\end{equation}
this can be written in the form
\begin{equation}
    p_i=\left( 1+(q_{ss}-1) (\alpha +\beta E_i)^\frac{1}{\delta}\right)^{-\frac{1}{q_{ss}-1}}.
\end{equation}
As a result of the entropy maximization, we thus obtain stretched $q$-exponentials as the generalized canonical distributions. The constant $\alpha$ needs to be adjusted in such a way that the normalization condition $\sum_{i=1}^Wp_i=1$ is satisfied. 

Note that the stretched form of the generalized canonical distribution is directly derived from the entropic functional
$S_{q,\delta}$ by employing the usual, physically plausible energy constraint
$\sum p_iE_i=const$. Alternatively, one could just maximize $S_q=S_{q,1}$, but in this case one would have to employ a more exotic stretched energy constraint when maximizing $S_q$, namely $\sum p_i E_i^\frac{1}{\delta}=const$. The use of $E_i^{1/\delta}$ rather than $E_i$ may be regarded as a `mesoscopic' area law \cite{Tsallis2026}. However, we do not need this non-standard constraint here, because in our approach the stretching follows directly from $S_{q,\delta}$ with the ordinary energy constraint. 


{\it Continuum case -} In the continuum case, we take $E_i=\frac{1}{2} u^2$, i.e.\ the fluctuating energy levels $E_i$ are given by the kinetic energy associated with a velocity difference $u$ in the turbulent flow at a given scale $r$. From this we get the prediction
\begin{equation}
    p(u)= \left( 1+(q_{ss}-1) \left( \alpha + \frac{1}{2} \beta u^2\right)^\frac{1}{\delta} \right)^{-\frac{1}{q_{ss}-1}}, \label{13}
\end{equation}
where $\alpha$ is to be determined from the normalization condition
\begin{equation}
    1= \int_{-\infty}^{+\infty} p(u)du.
\end{equation}
Physically, the reasoning for the occurrence of a stretching exponent $1/\delta$ in turbulent flows is sometimes given by the argument that the eddies in a turbulent flow are not space-filling.
To account for a slight skewness in the distributions, one may also introduce a small asymmetric correction term to the energy levels, see \cite{BLS} for details.

{\it Escort distributions -} In nonextensive statistical mechanics it is well-known that the escort distributions \cite{BS, Tsallis-Mendes-Plastino} defined as
\begin{equation}
    P_i = \frac{p_i^q}{\sum p_i^q}
\end{equation}
play an important role.

Given the stretched $q$-exponential form of eq.~(\ref{10}) we have the asymptotic behaviour (for large $E_i$)
\begin{equation}
    p_i\sim E_i^{\frac{1}{\delta}\frac{1}{q-1}}
\end{equation}
and we now see that for the special choice $\frac{1}{\delta}=q$ we get the asymptotic behaviour of an escort distribution
\begin{equation}
    p_i \sim E_i^{\frac{q}{q-1}}. \label{17}
    \end{equation}
Thus the stretching exponent $\frac{1}{\delta}$ can be eliminated if we choose $\frac{1}{\delta}=q$ and proceed to an escort theory with just one usual parameter $q$ (and no other parameter) and the non-stretched $E_i=\frac{1}{2}u^2$. In other words, eq.~(\ref{17}) can be re-interpreted as corresponding to an escort distribution $P_i$, but for ordinary (non-stretched) energy levels. This is 
possible {only} if
\begin{equation}
\frac{1}{\delta}=q=2-q_{ss}. \label{here}
\end{equation}
From the entropic point of view, we have thus (asymptotically) reduced the problem of a generalized statistical mechanics based on $S_{q,\delta}$ to an escort theory based on $S_q$ alone, which is possible only along the particular line $\delta=1/q=1/(2-q_{ss})$. 


{Having theoretically derived the relations (\ref{13}) and (\ref{here}), we are now in a position to compare with experimentally measured data. For this we use Taylor-Couette flow data measured by Swinney and co-workers that were previously empirically fitted \cite{BLS}.}
Note that in the paper \cite{BLS} the stretching exponent $\frac{1}{\delta}$ was denoted by $\alpha$, and $q_{ss}$ was denoted by the symbol $q$. It was empirically found in that paper that the relation (\ref{here}) is satisfied for turbulent Taylor-Couette flow data and yields {very good} fits for measured PDFs in the flow, see our Fig.~1 to illustrate this. But now, for the first time, we have derived the above relation between $q$ and $\delta$ 
{from a well-defined and useful principle in the nonextensive statistical mechanics formalism, namely the equivalence of a stretched theory with an escort theory. For the original introduction of escort distributions and previous uses of the escort formalism in $q$-statistics, we refer the reader to \cite{BS,Tsallis-Mendes-Plastino}}.

\begin{figure}[t]
\centering
\includegraphics[scale=0.43]{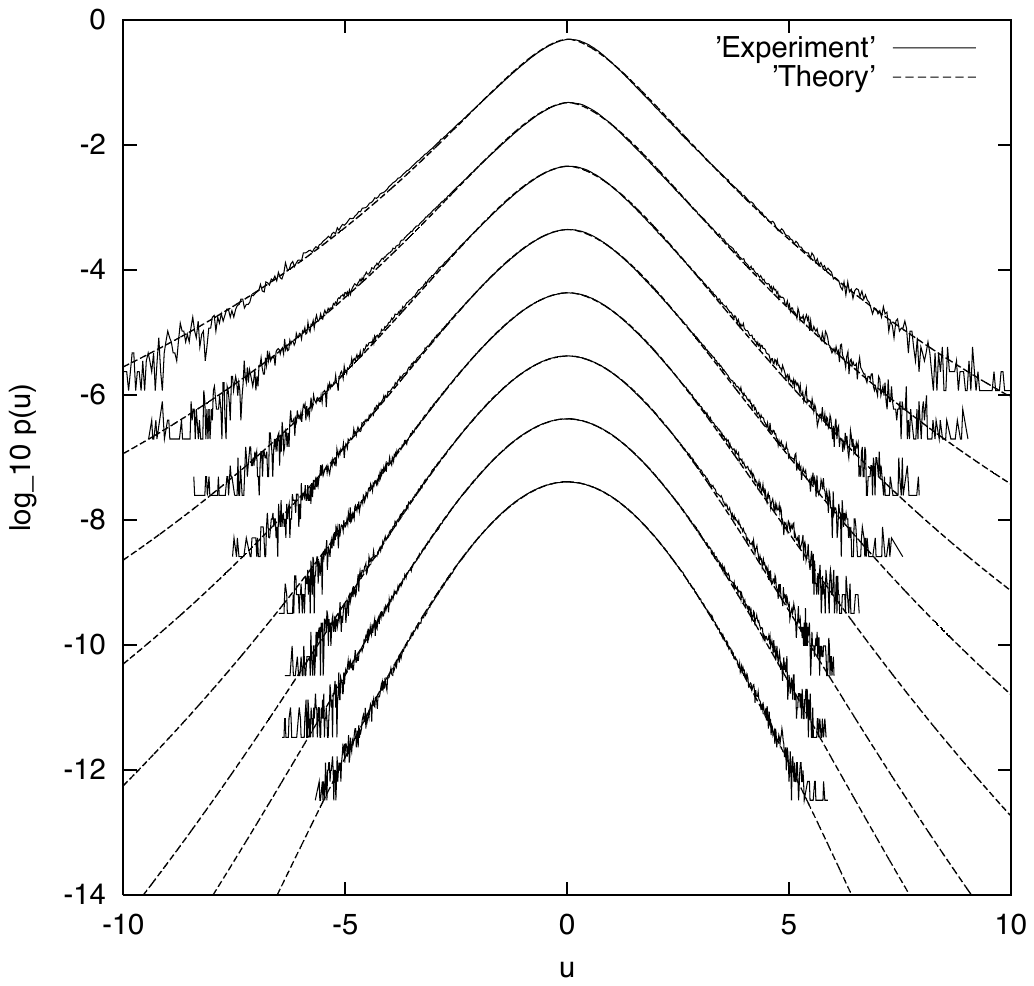}
\caption{Turbulent velocity distributions as measured in \cite{BLS} and a fit by the stretched $q$-exponentials, theoretically derived in the current paper from a maximum entropy principle for $S_{q,\delta}$.}

\label{Fig1a}
\end{figure}

{To further demonstrate the excellent agreement between our theoretically derived formulas and the experimental data, we also show a linear plot for the same data and fits in Fig.~2. This plot emphasizes the
vicinity of the maximum rather than the tails. In all cases, the disagreement  between data and theory is of the order of a few percent only.
We are not aware of any other competing analytic model that explains the shape of measured turbulent Taylor-Couette PDFs that well.}

\begin{figure}[t]
\centering
\includegraphics[scale=0.43]{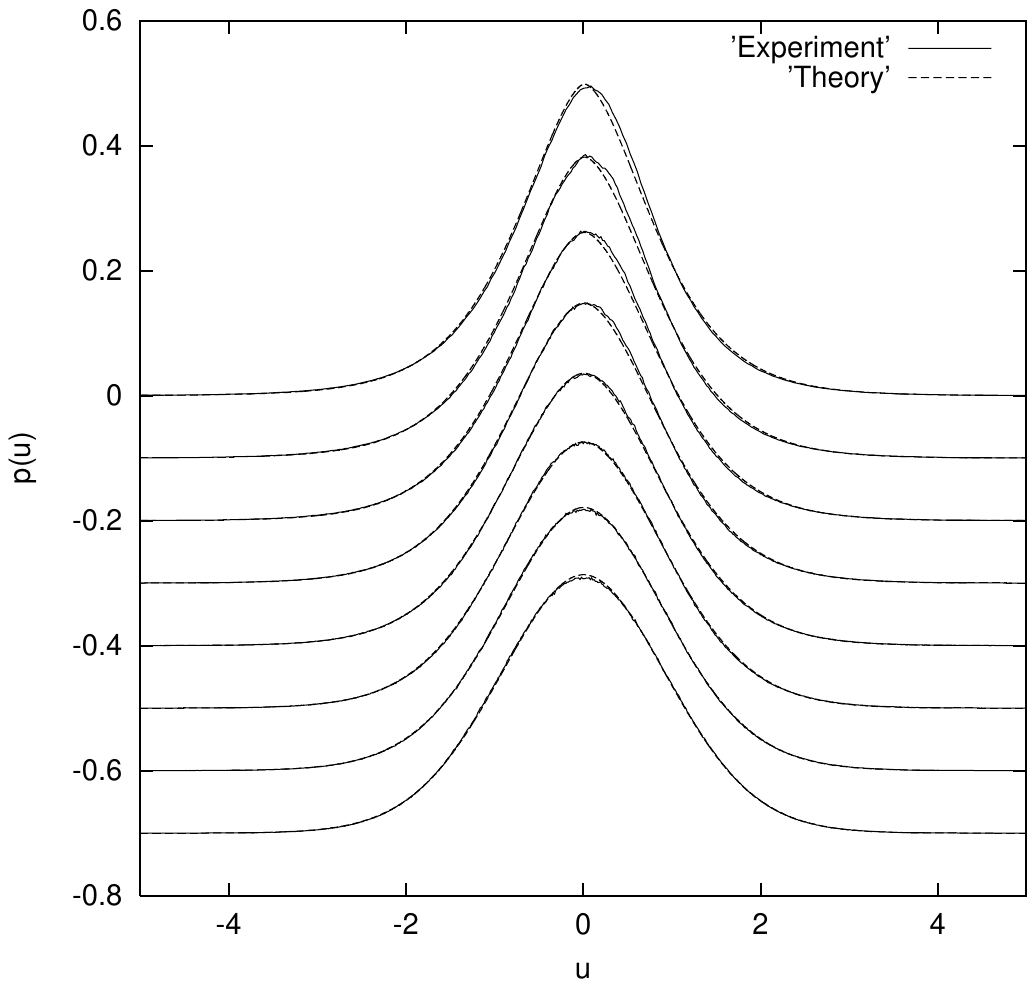}
\caption{{Same as Fig.~1, but in a linear plot. The largest deviation between data and theoretically predicted PDF is of the order of a few percent only, and observed on the right hand side of the maximum. Only one parameter is fitted, namely $q_{ss}$, while the {stretching exponent $\delta$} is given by eq.~(\ref{here}).}}

\label{Fig1b}
\end{figure}


{\it Physical interpretation of $q$ and $\delta$ -} In many physical applications of $q$-statistics, in
particular in nonequilibrium situations, an entropic exponent $q_{ss}>1$ is often associated with a superstatistics \cite{Beck2001,BeckCohen2003,BCS}, i.e.\ one considers driven nonequilibrium situations with local fluctuations of either the inverse temperature parameter $\beta$, or of the local energy dissipation rate in a turbulent flow, or of a general variance parameter. These
fluctuations, if occurring on a long time scale, generate $q$-statistics, either exactly or approximately. Generically one has the relation 
\begin{equation}
    q_{ss}= \frac {\langle \beta^2\rangle }{\langle \beta \rangle ^2}
\end{equation}
for type-A superstatistics, and slight modifications of it for type-B superstatistics \cite{BeckCohen2003,BCS,Beck2001}. 
The probability distribution $f(\beta)$ of the fluctuating parameter
$\beta$ can be a $\chi^2$ distribution, leading to exact $q$-statistics, but other probability densities such as e.g. the log-normal distribution or inverse $\chi^2$-distribution are allowed as well and have been considered in the literature \cite{BCS,beck2,SNC, Ourabah}. Superstatistics thus provides a plausible physical argument for the occurence of $q$-exponentials in many nonequilibrium situations \cite{peinke}.

The occurrence of a $\delta \not=1$ leading to stretched $q$-exponentials is a different effect and
is more of a geometric  nature. In \cite{Tsallis-Cirto,Tsallis2026} this is related to a dimensional argument for systems with an area-law anomaly, where
$\delta$ is given by
\begin{equation}
\delta =\frac{d}{d-1}
\end{equation}
in the microcanonical case, 
and $d$ denotes the dimension of the relevant physical space. The standard example is a black hole
where the standard Bekenstein-Hawking entropy lives on the 2-dimensional surface area of the black hole, therefore 
corresponding to dimension $d-1=2$,
whereas a thermodynamic entropy requires $d=3$, thus implying $\delta=3/2$ \cite{Tsallis-Cirto}.  The Barrow entropy, introduced in \cite{Barrow2020},  proposes values of $\delta$ that are slightly larger than 1, physically motivated due to an assumed fractal structure of the black hole surface. Canonical ensembles based on $S_{q,\delta}$ have been shown to have a $\delta$ inbetween 1 and the microcanonical case $3/2$ if $d=3$ \cite{Tsallis2026}.

For the new turbulence application described here, we indeed have $1 \leq \delta \leq \frac{3}{2}$,
depending on scale. An effective $\delta \not=1$ is a standard assumption in (multi-)fractal turbulence models, due to the fact that eddies in the turbulent flow are generally thought to be not space-filling. 
The endpoint of our turbulent cascading scenario will correspond to $\delta=3/2$ (hence $q_{ss}=4/3$) at the Kolmogorov scale $\eta$,
where turbulent structures will completely disappear, in certain analogy to a black hole that swallows the remaining turbulent structures at this scale.


{\it Experimental verification -} Let us now compare our theoretical predictions in more detail with experimental measurements. We put particular emphasis on checking the parameter dependencies that we have derived from theory. We refer to the results of the experiment described in \cite{BLS} for turbulent
Taylor-Couette flow. We consider a distance-dependent $q_{ss}=q_{ss}(r)$ for velocity differences measured between
two points in the liquid separated by a distance $r$,
and present the best fits of the measured PDFs. The parameter $q_{ss}$ decreases with increasing scale and
at largest scales ($r/\eta > 1000$) we have $q_{ss} \approx 1$ (apart from finite-size effects), as shown in Fig.~3. 

\begin{figure}[t]
\centering
\includegraphics[scale=0.57]{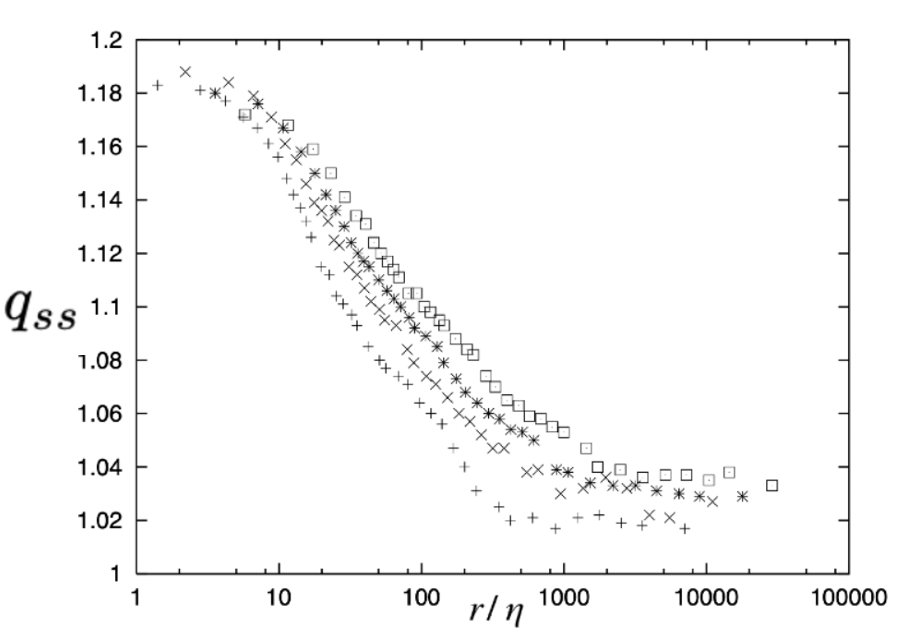}
\caption{$q_{ss}$ as a function of $r/\eta$ and $Re$ as measured in \cite{BLS}. $\delta$ is given by $\delta=(2-q_{ss})^{-1}$. The different symbols represent 4 different Reynolds numbers, namely $Re=69000, 133000, 266000, 540000$ (from left to right). 
}
\label{Fig3}
\end{figure}

At the smallest scales $(r=\eta)$ we have the analogue of a `black hole', i.e. the turbulent eddy structures disappear, and everything completely dissipates. When looking at the asymptotic decay rate of the measured PDFs using the data of \cite{BLS}, we observe excellent fittings of the asymptotic form
\begin{equation}
    p(u) \sim |u|^{-w}
\end{equation}
with
\begin{equation}
    w(r)= \frac{1}{\delta} \frac{2}{q_{ss}-1} 
    \end{equation}
where
\begin{equation}
    \frac{1}{\delta} =2-q_{ss} \label{constraint}
\end{equation}
and
\begin{equation}
    w(r)=4 \left(\frac{r}{\eta}\right)^\kappa .\label{24}
\end{equation}
This is illustrated in Fig.~4. The exponent $\kappa$ depends weakly on the Reynolds number, taking values  between 0.33 (largest Reynolds number in the experiment) and 0.44 (smallest Reynolds number in the experiment). For $r=\eta$, eq.~(\ref{24}) yields $w=4$. Note that this limit cannot be reached in the experiment but is extrapolated. 

{To find the optimum fitting parameter $q_{ss}$ for each experimentally measured distribution at a given scale $r$, we used standard statistical methods based on maximum likelihood estimates (MLE), incorporating the normalization condition and a rescaling to variance 1 together with the theoretically derived constraint (\ref{constraint}). 
Alternatively, a least square optimization can be used, yielding  the same results for the optimum $q_{ss}$ within the numerical precision.}

\begin{figure}[t]
\centering
\includegraphics[scale=0.38]{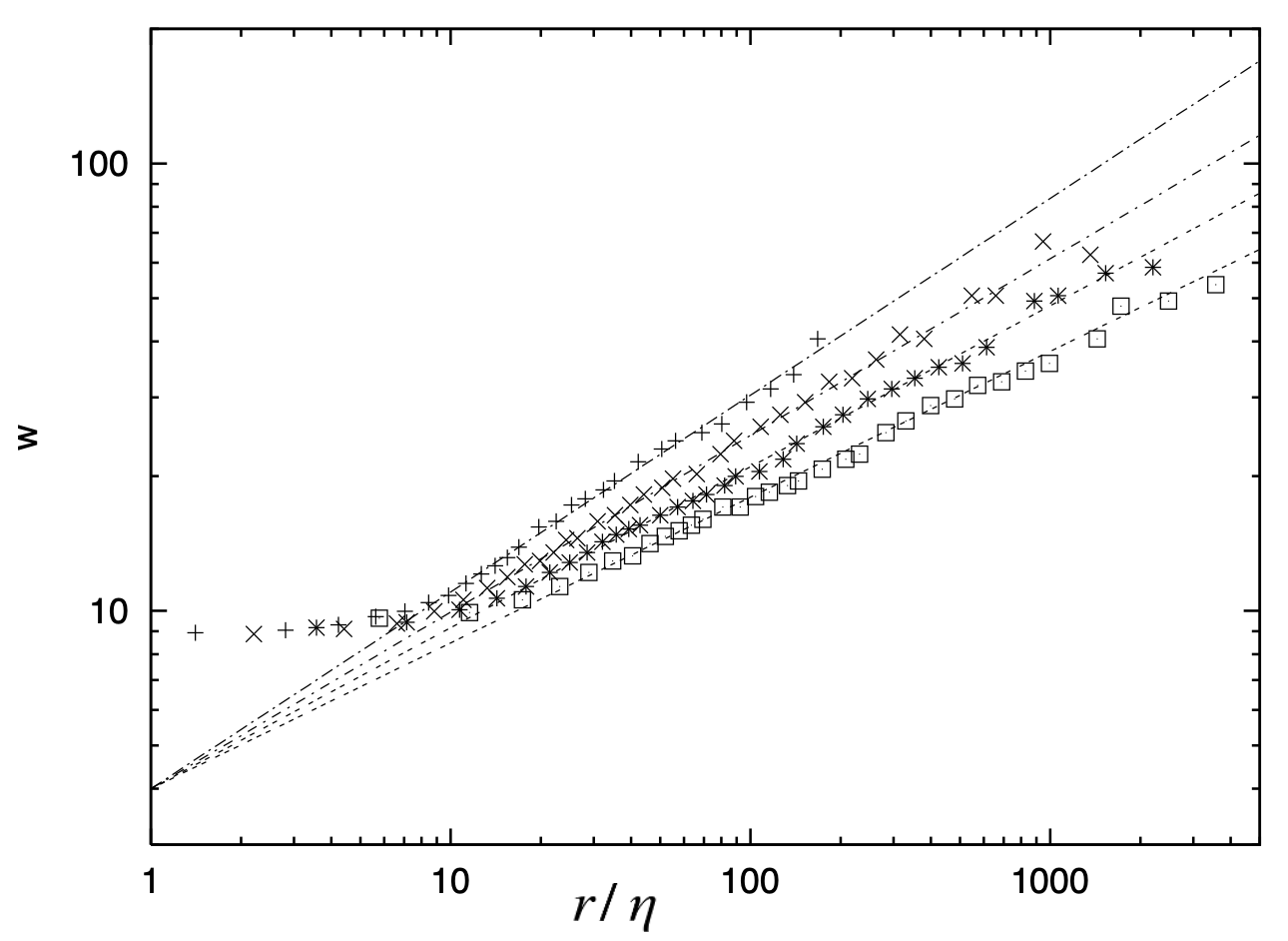}
\caption{The asymptotic power-law decay rate exponent $w$ of the velocity distributions as a function of $r/\eta$ for various Reynolds numbers, as measured in \cite{BLS}. A scaling law is observed which, if extrapolated to $r=\eta$, gives $w=4$ at the Kolmogorov scale $r=\eta$. 
}
\label{Fig-4}
\end{figure}
We may actually fit the entire data in Fig.~\ref{Fig-4} with the following functional form:
\begin{equation}
w(r) = w_0 [1 + (4/w_0)^{1/\kappa} r/\eta]^\kappa \,,
\end{equation}
where $w_0 \simeq 9$. This equation is an effective equation showing that the Kolmogorov scale can never be
reached in realistic turbulence experiments. It describes finite-size effects, providing us with the realistically achievable exponent $w$ at small scales.

The above measurements and fits of observed PDFs provide a perfect match to our entropic turbulence theory. We may write
\begin{equation}  
w(r)= 2 \cdot \frac{2-q_{ss}}{q_{ss}-1}
\end{equation}
for the asymptotic decay rate of the PDFs on the curve $\delta^{-1}=2-q_{ss}$, where we have equivalence to an escort theory based on $S_q$ on its own. Since $S_q$ on its own has particularly good properties, namely, it is trace-form and composable (and is in fact the unique entropic functional with these properties \cite{Tempesta2017}),
this means a well-defined thermodynamic description is possible on this particular curve \cite{Jensen-Jizba-Tempesta}. The turbulent flow appears to choose precisely these parameters in its scale-dependent behaviour.

Two limit cases are of particular interest:

\begin{itemize}
    
\item
$\frac{r}{\eta} \to \infty \Rightarrow w \to \infty \Rightarrow q_{ss} \to 1$ (hence $q\to 1$). On large scales we recover Boltzmann-Gibbs statistical mechanics.

\item
$\frac{r}{\eta} \to 1 \Rightarrow w \to 4 \Leftrightarrow q_{ss} \to \frac{4}{3}$. 
This means that if we approach the Kolmogorov scale, then $\delta =\frac{1}{2-q_{ss}}\to \frac{3}{2}$. This is 
precisely the value that, in a different context, is relevant for black hole physics and cosmology \cite{Tsallis-Cirto, jizba2022, TsallisJensen2025}.  
\end{itemize}

{The central result of our paper, namely the relation $\delta=1/(2-q_{ss})$, is plotted as the black curve in Fig.~5. Our main result is that turbulent Taylor-Couette flows don't explore the entire parameter space, but prefer the black line.}

\begin{figure}[t]
\centering
\includegraphics[scale=0.4]{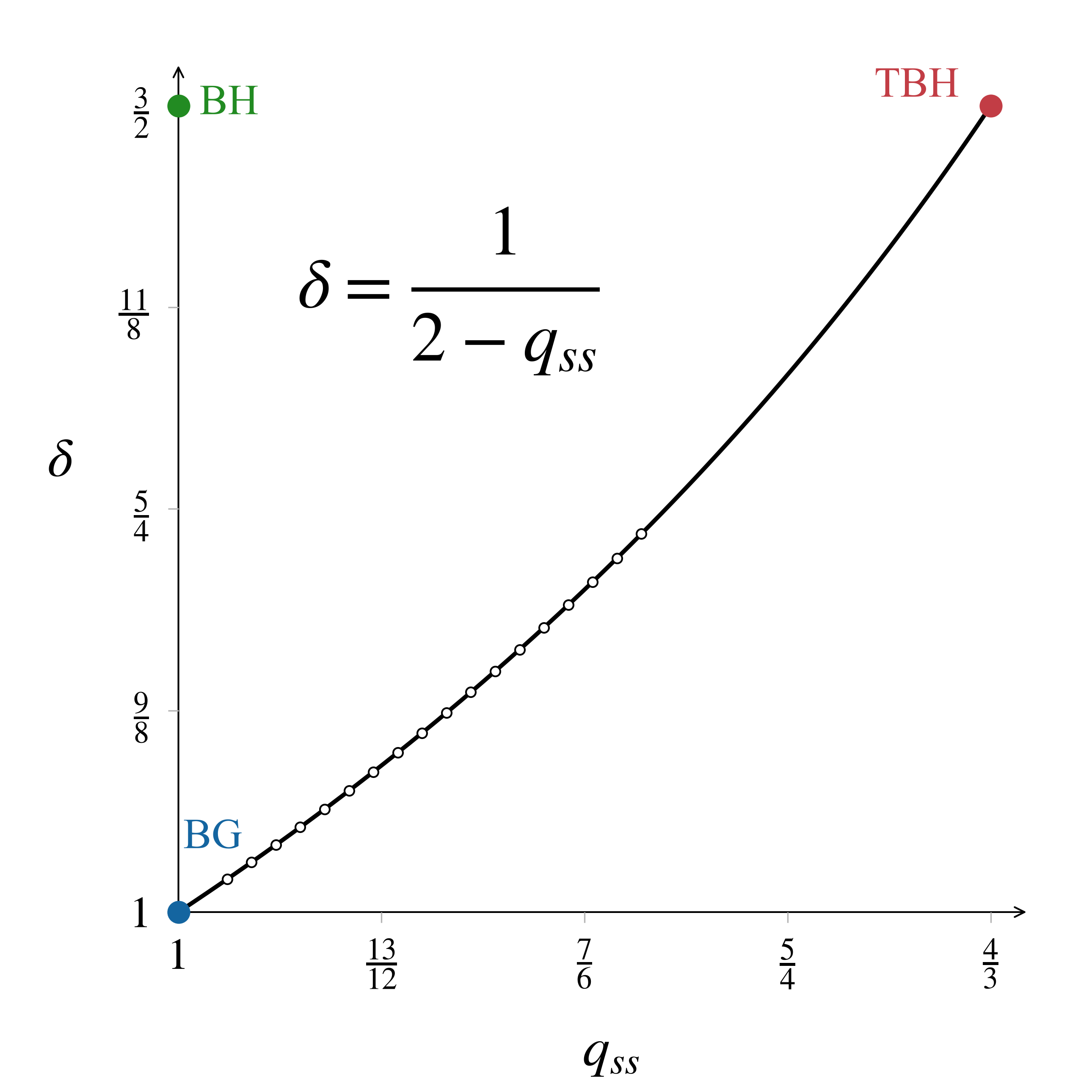}
\caption{Visualization of the main result of this paper. The point (1,1) corresponds to ordinary Boltzmann-Gibbs (BG) statistical mechanics. The point $(1, \frac{3}{2})$ has been used in previous papers \cite{Tsallis-Cirto, TsallisJensen2025} for a generalized thermodynamic description of black hole (BH) physics. The turbulent Taylor-Couette flow, as illustrated in this paper for data measured in \cite{BLS}, chooses its parameters not in the full 2-dimensional $(q_{ss},\delta)$ plane but along the black line given by $\delta=(2-q_{ss})^{-1}$ (data points in the region  $1.02 \leq q_{ss}\leq  1.19$ as shown in Fig.~3). The end point of the turbulent cascade is given by $(\frac{4}{3}, \frac{3}{2})$, at this point the 3rd moment of velocity differences ceases to exist. Formally, this end point (which is not reachable in the experiment) corresponds to the entropic parameters of a turbulent black hole (TBH) since $\delta=\frac{3}{2}$.} 

\end{figure}

Because $w=4$ at the smallest scale $r=\eta$, this means the third moment $\langle u^3 \rangle$ (which is very important in any turbulence model) ceases to exist. Overall, we have shown that our entropic turbulence theory is
experimentally very well confirmed by the data measured in \cite{BLS}.
It means that at the Kolmogorov scale,
we have, in the entropic sense, a situation where all turbulent eddy structures completely disappear and everything dissipates. This has a certain analogy with a black hole-type of state absorbing all measurable structures
at the endpoint of the cascade.


{\it Conclusion and outlook-} We presented a new entropic approach to turbulence that involves a generalized entropic functional $S_{q,\delta}$
which is maximized subject to suitable constraints,
providing a testable statistical mechanics formalism for turbulent flow. The entropy involved, $S_{q,\delta}$,
has been previously shown to have applications in cosmological models \cite{jizba2024,luciano2025,TsallisJensen2025}.
Our main result is a derived relation between $q$ and $\delta$ relevant for turbulent flow.
We described a formal analogy between our scale-dependent entropic turbulence theory and a kind of black hole state that is (in a statistical sense) absorbing the eddies at the Kolmogorov scale.  It should be clear that this is a formal analogy only. Nevertheless, the turbulence aspects of our theory are experimentally confirmed, in the sense that the maximization of $S_{q,\delta}$ yields the statistical properties of the observed PDFs in Taylor-Couette flow as measured in \cite{BLS}, and provides measured evidence for the relevance of the parameter $\delta=3/2$ at the end point of the cascade. The value $\delta =3/2$ was previously proposed to describe a thermodynamic entropy for black hole states \cite{Tsallis-Cirto}.

Whether our new entropic turbulence theory just involves a formal black-hole analogy or also describes turbulence close to true gravitational black holes, 
is an open question worth further investigation. In \cite{Yang-et-al} it is shown that rapidly spinning astrophysical black holes trigger turbulent behaviour in their surrounding environment. In the context of our paper here, such a rapidly spinning black hole has certain similarities with a kind of `gravitational Taylor-Couette experiment'. If suitable astrophysical data were available for turbulent matter close to astrophysical black holes one could test our predictions for the shape of the PDFs. This question is certainly worthwhile to investigate in future research, in particular in view of recent experimental observations of turbulence near supermassive black holes \cite{nature2026,xrism2025}. 

{\it Acknowledgements - }
We have benefited from  useful remarks from H.S. Lima. C.B. acknowledges
funding from an institutional QMUL ISPF-ODA grant allocated to QMUL by Research England,
as well as from the STFC grant UKRI467. C.T. acknowledges partial financial support from CNPq and Faperj (Brazilian agencies). {We dedicate this paper to the memory of Harry L. Swinney (1939-2026) whose group measured the relevant Taylor-Couette data \cite{BLS} that were analysed here}.


\end{document}